\documentclass[aps,prl,reprint,superscriptaddress]{revtex4-1}
\usepackage{graphicx}
\usepackage{bm}
\usepackage{color}
\usepackage{amsmath}
\usepackage{amsfonts}
\usepackage{amssymb}
\usepackage{natbib}
\usepackage{latexsym}
\usepackage{float}
\usepackage{verbatim}

\begin{document}

\title{Pitch Angle Anisotropy Controls Particle Acceleration and Cooling in \\ Radiative Relativistic Plasma Turbulence}
\author{Luca Comisso}
\email{luca.comisso@columbia.edu}
\affiliation{Department of Astronomy and Columbia Astrophysics Laboratory, Columbia University, New York, NY 10027, USA}
\author{Lorenzo Sironi}
\affiliation{Department of Astronomy and Columbia Astrophysics Laboratory, Columbia University, New York, NY 10027, USA}

\begin{abstract} 
Nature's most powerful high-energy sources are capable of accelerating particles to high energy and radiate it away on extremely short timescales, even shorter than the light crossing time of the system. It is yet unclear what physical processes can produce such an efficient acceleration,  despite the copious radiative losses. By means of radiative particle-in-cell simulations,  we show that magnetically dominated turbulence in pair plasmas subject to strong synchrotron cooling generates a nonthermal particle spectrum with a hard power-law range (slope $p \sim 1$) within a few eddy turnover times. Low pitch-angle particles can significantly exceed the nominal radiation-reaction limit,  before abruptly cooling down. The particle spectrum becomes even harder ($p < 1$) over time owing to particle cooling with an energy-dependent pitch-angle anisotropy. The resulting synchrotron spectrum is hard ($\nu F_\nu \propto \nu^s$ with $s \sim 1$). Our findings have important implications for understanding the nonthermal emission from high-energy astrophysical sources,  most notably the prompt phase of gamma-ray bursts and gamma-ray flares from the Crab nebula.
\end{abstract}

\maketitle

A variety of astrophysical sources,  ranging from pulsar wind nebulae (PWNs) to gamma-ray bursts (GRBs) and active galactic nuclei (AGNs),  are capable of accelerating particles to gamma-ray-emitting energies \cite{Buhler2014,Meszaros2006,Madejski2016}.  The high radiation efficiency of these sources requires that the energy transferred to the particles must be carried away by radiation on extremely short timescales,  comparable or even shorter than the light-crossing time of the system \cite{Buhler2014,Meszaros2006,Madejski2016}.  Under these circumstances,  the physical mechanism responsible for particle acceleration has to compete with significant radiative losses, and the physical origin of the observed emission remains unclear.

Dissipation of the large reservoir of magnetic energy in the aforementioned systems \cite{Porth17,MacFadyen99,Ruiz19,Blandford2019} provides a conceivable path toward particle acceleration and copious radiative emission.  In view of the enormous scale separation between the energy-carrying scale and the plasma kinetic scales,  turbulence is a natural candidate for converting the available magnetic energy into particle kinetic energy \cite{Biskamp03,FP2008}.  In highly magnetized turbulent plasmas with negligible radiative losses,  the interplay of magnetic reconnection  \cite{ComissoSironi18,ComissoSironi19,Comisso20,NB2021} and stochastic scattering off turbulent fluctuations \cite{ComissoSironi18,Zhda2018,ComissoSironi19,Wong2020} has been shown to produce high energy particles with robust power-law distributions. However,  power-law spectra might be suppressed or steepened by strong radiative cooling, as observed in simulations of turbulence with inverse Compton scattering off external photons \cite{Zhda2020,Sobacchi2021a,Zhda2021,NB2021}.

Numerous astrophysical sources \citep[e.g.][]{Abdo11,Burgess20,Tavecchio10} 
radiate a large fraction of energy via the synchrotron mechanism,  which is usually the main radiative channel in magnetically dominated plasmas. Under these conditions,  heuristic arguments suggested that turbulence would produce a quasi-thermal electron energy distribution \cite{Uzdensky2018} or a steep ($p \geq 2$) nonthermal power law \cite{Sobacchi2021b},  but no first-principles simulations have been conducted so far. Fully kinetic simulations are necessary in order to capture the interplay between particle acceleration,  scattering,  and cooling, allowing to self-consistently determine the resulting particle distribution. An assessment that takes into account both energy and pitch angle of the particles is essential for determining the resulting synchrotron emission,  and is required to obtain falsifiable,  predictive models of astrophysical high energy sources.

In this paper,  we adopt a first principles approach by solving the kinetic plasma equations through the particle-in-cell (PIC) method \cite{birdsall85} using the PIC code TRISTAN-MP \cite{buneman93, spitkovsky05}.  We take into account the emission of radiation by the particles via the inclusion of the radiation reaction force $\bm{F}_{RR}$ in the particle equation of motion,  which,  in the reduced Landau-Lifshitz form,  is given by \cite{Vranic16}
\begin{eqnarray}\label{F_RR}
\bm{F}_{RR} &=&\frac{2}{3} r_0^2 \Big[(\bm{E} + \bm{\beta} \times \bm{B}) \times \bm{B} + (\bm{\beta} \cdot \bm{E})\bm{E} \Big] \nonumber\\
&& - \frac{2}{3} r_0^2 \gamma^2 \bm{\beta} \Big[ (\bm{E} + \bm{\beta} \times \bm{B})^2 -(\bm{\beta} \cdot \bm{E})^2 \Big] \, , \; \;
\end{eqnarray}
where ${\bm{\beta}}$ indicates the dimensionless velocity of the particle,  $\gamma$ is its Lorentz factor,  ${\bm{E}}$ and ${\bm{B}}$ are the electric and magnetic fields,  and $r_0 = e^2/m_e c^2$ stands for the classical electron radius.  This expression for $\bm{F}_{RR}$ gives accurate results for parameters of interest where the classical description of the particle motion is applicable,  i.e.  $\gamma B/B_{\rm QED} \ll 1$, with $B_{\rm QED}=m_e^2 c^4/e \hbar \simeq 4.4 \times 10^{13} {\rm \; G}$.

We initialize a uniform electron-positron plasma with total particle density $n_0$ according to a Maxwell-J\"{u}ttner distribution ${f_0}(\gamma) = ({\gamma^2 \beta}/{\theta_0}{K_2}(1/{\theta_0})) \, e^{- {\gamma}/\theta_0}$ with dimensionless temperature $\theta_0 = {k_B T_0}/{m_e c^2} = 0.3$.  Here,  $T_0$ is the initial plasma temperature,  $k_B$ indicates the Boltzmann constant,  and $K_2(x)$ is the modified Bessel function of second kind. The corresponding mean particle Lorentz factor is $\gamma_{th0} \simeq 1.6$. Turbulence develops from uncorrelated magnetic field fluctuations initialized in the plane perpendicular to a uniform mean magnetic field  $\langle{\bm{B}}\rangle=B_0{\bm{\hat z}}$. The initial fluctuations have low wavenumbers $k_j = 2\pi n_j/L$,  with $n_j \in \{ {1, \ldots ,4} \}$ and $j=x,y,z$.  Setting equal amplitude per mode, the initial magnetic energy spectrum peaks near $k_p = 8 \pi /L$, which defines the energy-carrying scale $l = 2 \pi /k_p$.

To capture the full turbulent cascade from macroscopic/fluid scales to kinetic scales, we adopt a periodic cubic box of volume $L^3$ consisting of $3072^3$ cells of size $\Delta x = d_{e0}/3$,  where $d_{e0}=c/\omega_{p0}$ indicates the initial plasma skin depth and $\omega_{p0} =\sqrt {4\pi n_0 {e^2}/\gamma_{th0} {m}}$ is the relativistic plasma frequency.  We employ an average of 8 computational particles per cell. We have verified with smaller 3D simulations that the discussed results are the same when using up to 128 particles per cell.

The strength of initial fluctuating magnetic energy relative to plasma enthalpy is quantified by $\sigma_{\delta B}= {\delta B_{{\rm{rms}}0}^2}/{4\pi n_0 w_0 m c^2}$,  where $w_0 = \gamma_{th0} + \theta_0$ is the enthalpy per particle and $\delta B_{{\rm{rms}}0} = \langle {\delta {B^2} (t=0)} \rangle^{1/2}$. We consider strong turbulence with $\delta B_{{\rm{rms}}0} = B_0$.  The corresponding total magnetization is $\sigma  = \sigma_{B_0} + \sigma_{\delta B}$,  with $\sigma_{B_0}  = { B_0^2}/{4\pi n_0 w_0 m c^2}$.  Since we are interested in the relativistic regime, we take $\sigma_{B_0} = \sigma_{\delta B} = 50$,  which yields the Alfv{\'e}n speed ${v_{A}} = c\sqrt {{\sigma}/(1 + {\sigma})}  \simeq c$.

The strength of the radiation reaction force is parametrized by the Lorentz factor ($\gamma_{\rm{rad}}$) for which the drag force balances the accelerating force. For ultrarelativistic particles ($\gamma \gg 1, \, \beta \simeq 1$),  the balance between the accelerating electric force, $F_{\rm{acc}} = e E$, and the drag force induced by synchrotron losses, $F_{RR}^{\rm{sync}}  \simeq  (2/3) r_0^2 \gamma^2  B^2 \sin^2 \alpha$ (where $B$ is a fiducial magnetic field and $\alpha$ is the pitch angle evaluated in the fluid frame),  gives 
\begin{equation} \label{gamma_rad}
\gamma_{\rm{rad}} = \frac{1}{\sin \alpha} \sqrt{\frac{3 \beta_E}{2} \frac{B_{\rm{cl}}}{B}} \, ,
\end{equation}
where $B_{\rm{cl}} = e/r_0^2 = m_e^2 c^4/e^3$ is the critical classical magnetic field strength and  $\beta_E = E/B$ is the ratio between the fiducial electric and magnetic fields.

\begin{figure}
\begin{center}
\hspace*{-0.010cm}
\includegraphics[width=8.75cm]{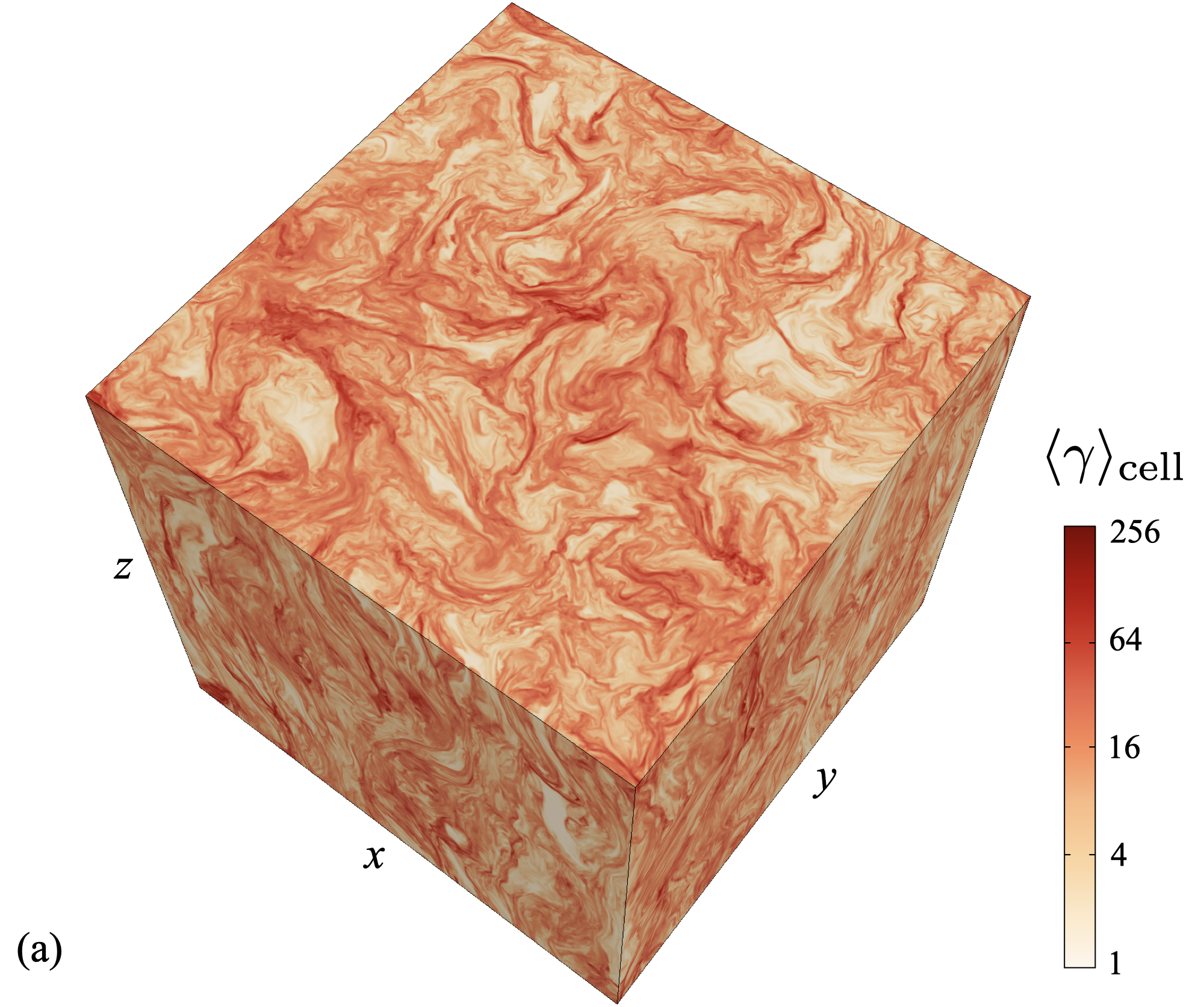} \\
\vspace*{0.3cm}
\includegraphics[width=8.75cm]{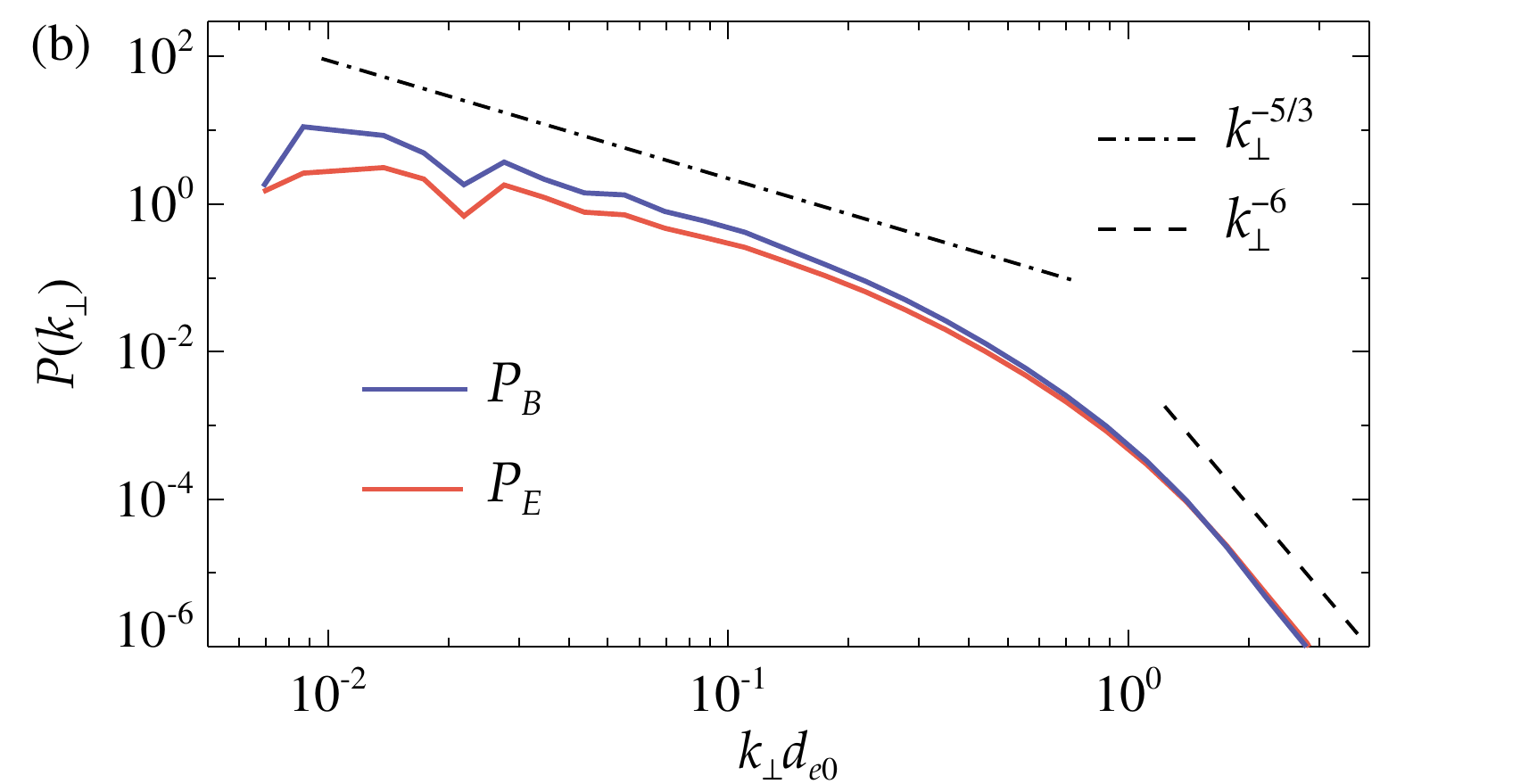}
\end{center}
\vspace{-0.4cm}
\caption{(a) 3D plot of $\langle \gamma \rangle_{\rm{cell}}$ from a simulation with $\gamma_{\rm{rad}}=75$ at $t \sim 2.5 l/c$,  indicating that particle acceleration is highly localized in space.  (b) Power spectra for the turbulent magnetic field (blue) and the electric field (red) at $t \sim 2.5 l/c$.  Slopes $k_\bot^{-5/3}$ (dotted-dashed line) and $k_\bot^{-6}$ (dashed line) are shown for reference. }
\label{fig1}
\end{figure}

Particles are strongly cooled if they radiate a significant fraction of their energy in a timescale shorter than the magnetic field dissipation timescale.  Therefore,  it is convenient to define the Lorentz factor $\gamma_{\rm{cool}}$ of a particle that cools in a few ($\kappa$) outer-scale eddy turnover times,  i.e.  $\tau_{\rm{cool}}(\gamma_{\rm{cool}})= \kappa l/c$.  From the synchrotron cooling timescale $\tau_{\rm{cool}} = {\gamma}/{{\left| {d\gamma/dt} \right|}} = {{3 {m_e}{c}}}/({{2{r_0^2} \gamma {B^2} \sin^2 \alpha}})$, one can express $\gamma_{\rm{cool}}$ in terms of $\gamma_{\rm{rad}}$ as
\begin{equation} \label{gamma_cool}
\gamma_{\rm{cool}} = \frac{c}{\kappa l}  \,  \frac{\gamma_{\rm{rad}}^2}{\omega_L \beta_E}  = \frac{1}{\sqrt{\sigma_{B_0} \gamma_{th0} w_0}} \frac{d_{e0}}{l} \,   \frac{\gamma_{\rm{rad}}^2}{\kappa \beta_E}  \, ,
\end{equation}
where $\omega_L=eB_0/mc$ is the nonrelativistic Larmor frequency. In this Letter,  we are interested in the regime where most particles are rapidly cooled,  which is given by the hierarchy $\gamma_{\rm{cool}} <  \gamma_{\sigma} < \gamma_{\rm{rad}} < \gamma_{\rm max}$,  where $\gamma_{\rm max} \sim e B \, l/m c^2 \sim \sqrt{\sigma_{B_0}} \gamma_{th0} (l/d_{e0})$ is the Lorentz factor at which the particle Larmor radius reaches the outer-eddy scale, while $\gamma_\sigma = \gamma_{th0} + \sigma_{\delta B} w_0/2 \simeq 48.5$ is the mean particle Lorentz factor assuming complete turbulent field dissipation. Therefore,  here we present results from simulations with $\gamma_{\rm{rad}} \in \left\{ {75,125,200,\infty} \right\}$,  where, in Eq. \eqref{gamma_rad},  we used $E = 0.1 \delta B_{{\rm{rms}}0} $ as appropriate for fast magnetic reconnection \cite{ComissoJPP16,Liu17},  and ${\sin^2 \alpha} = 2/3$ as nominal value for an isotropic pitch-angle distribution.  Hence,  $\gamma_{\rm{cool}}(\kappa=1) \in \left\{ {18,50,128,\infty} \right\}$, which allows us to compare different cooling regimes.

\begin{figure}
\begin{center}
\includegraphics[width=8.75cm]{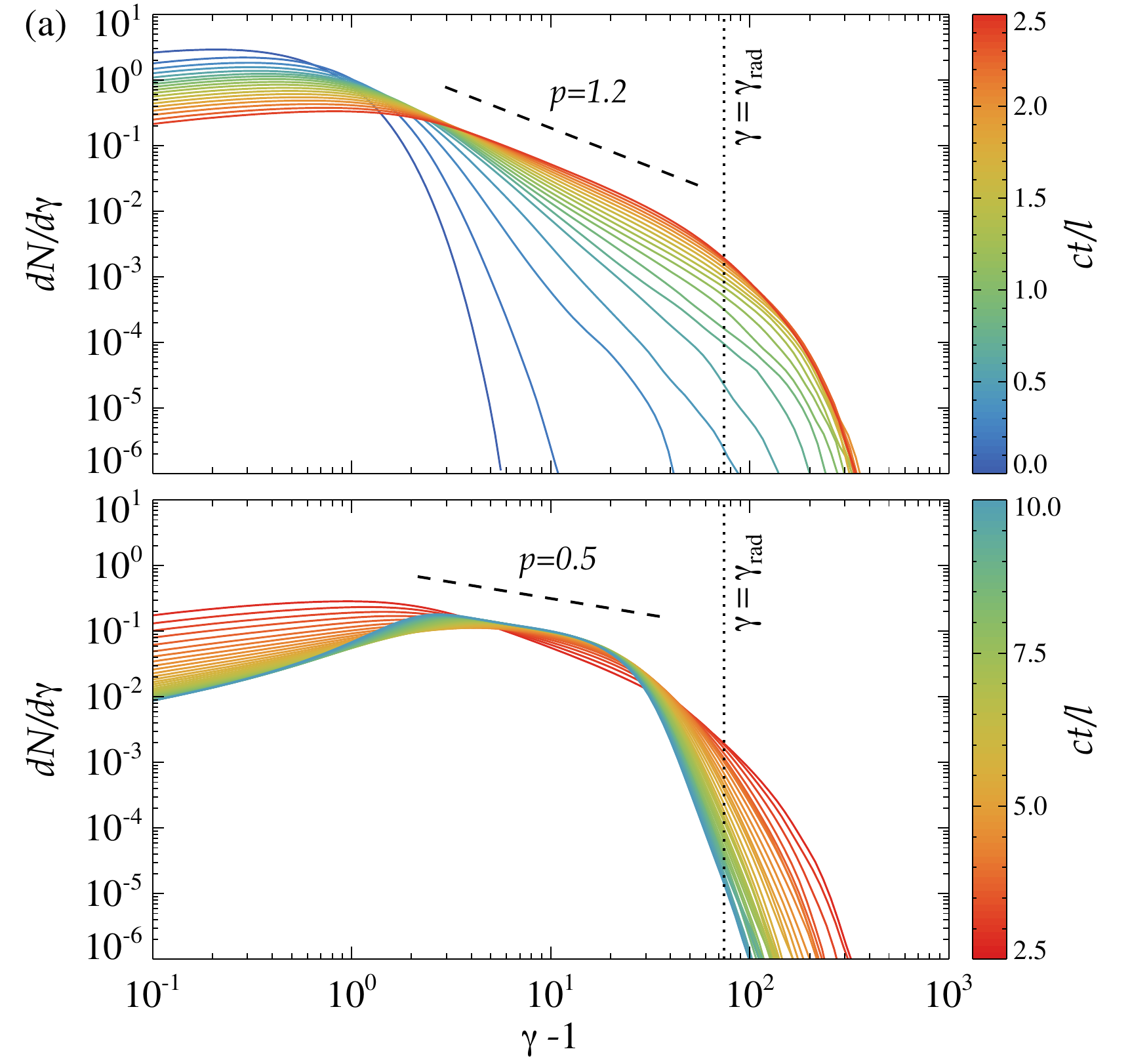}
\includegraphics[width=8.75cm]{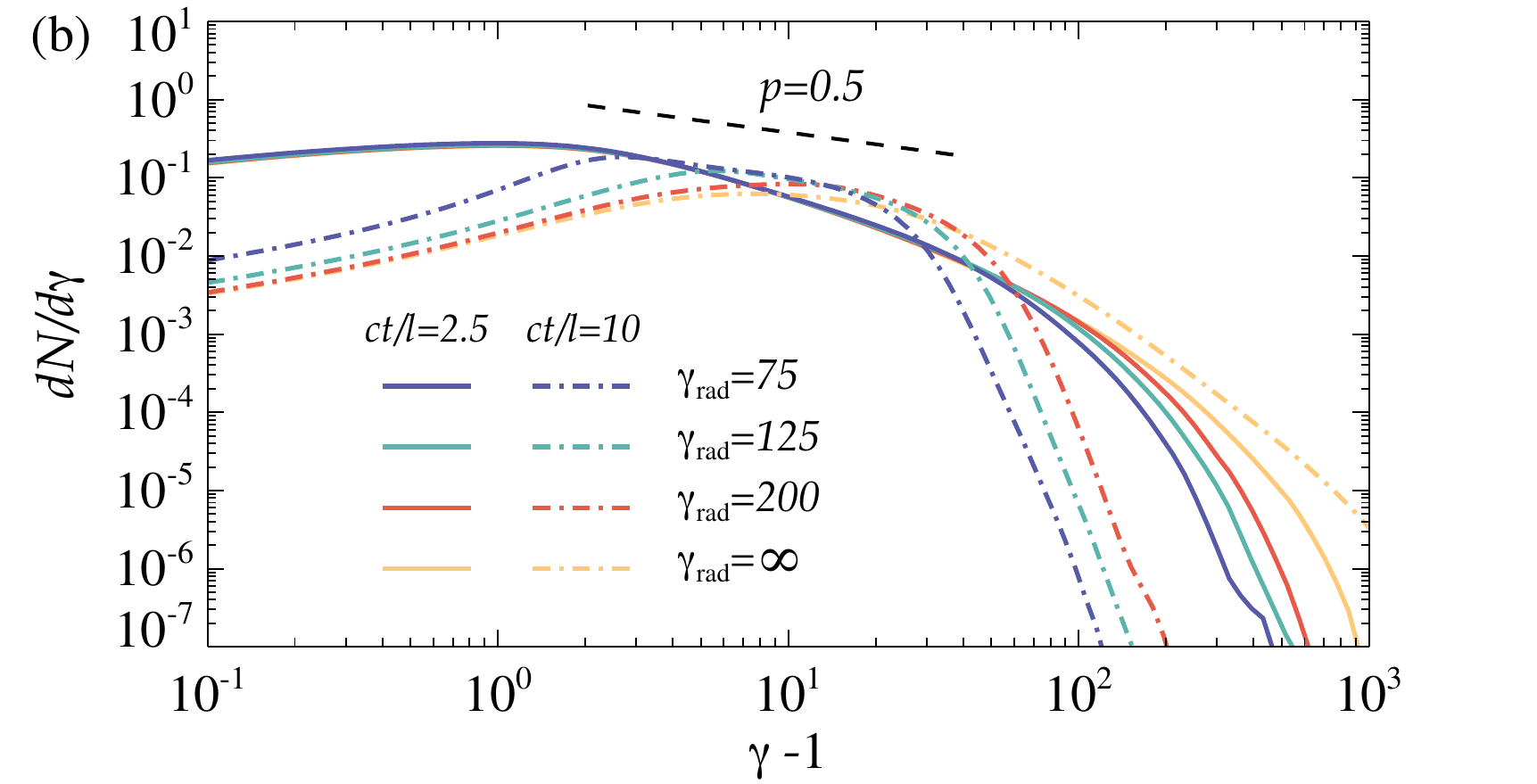}
\end{center}
\vspace{-0.4cm}
\caption{(a) Time evolution of the particle energy distribution separated in two time intervals for the simulation in Fig.~\ref{fig1}.  (b) Particle energy distributions at $ct/l=2.5$ (solid lines) and $ct/l=10$ (dotted-dashed lines) for simulations with $\gamma_{\rm{rad}} \in \left\{ {75,125,200,\infty} \right\}$.  }
\label{fig2}
\end{figure}

\begin{figure}[]
\begin{center}
\includegraphics[width=8.75cm]{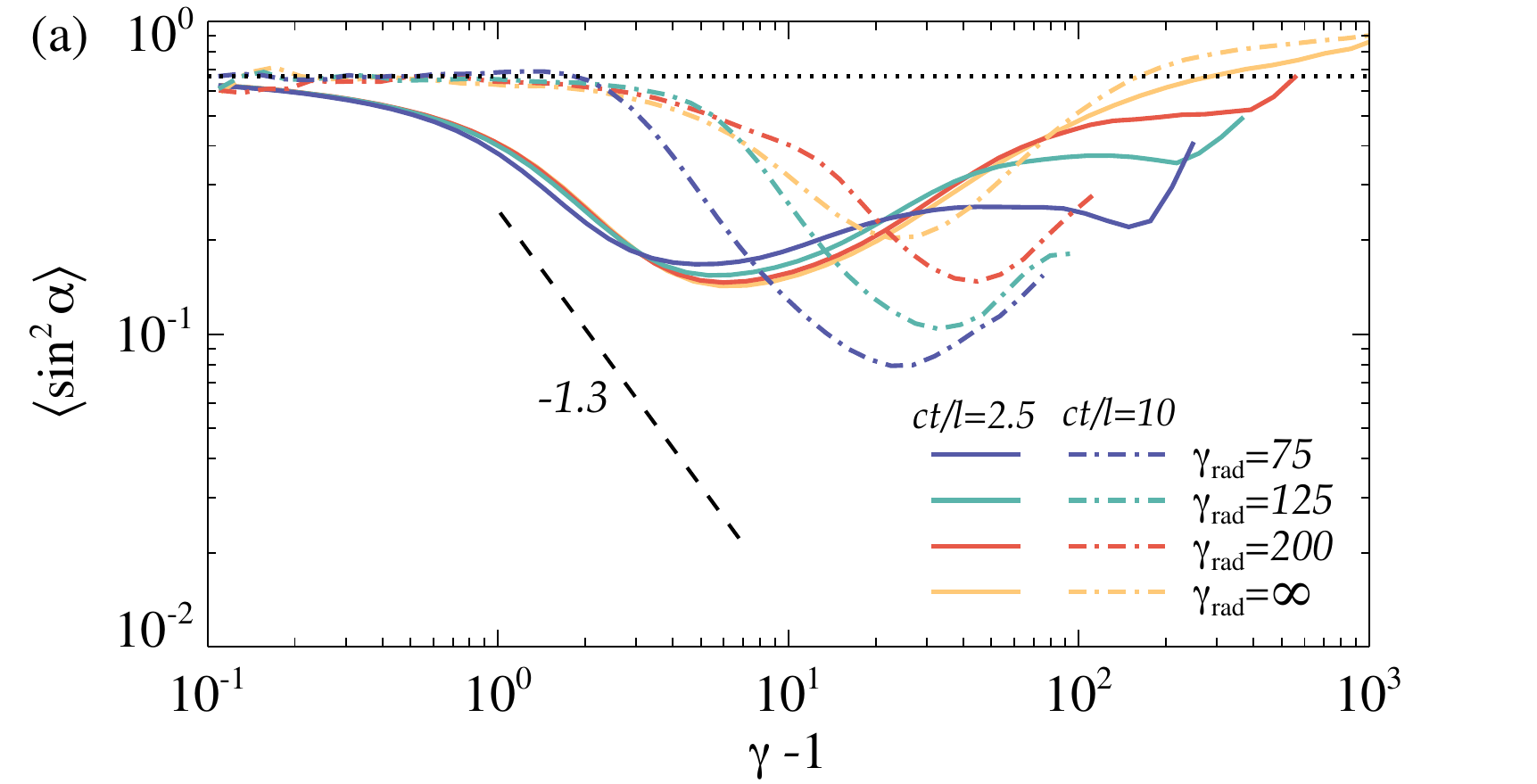}
\includegraphics[width=8.75cm]{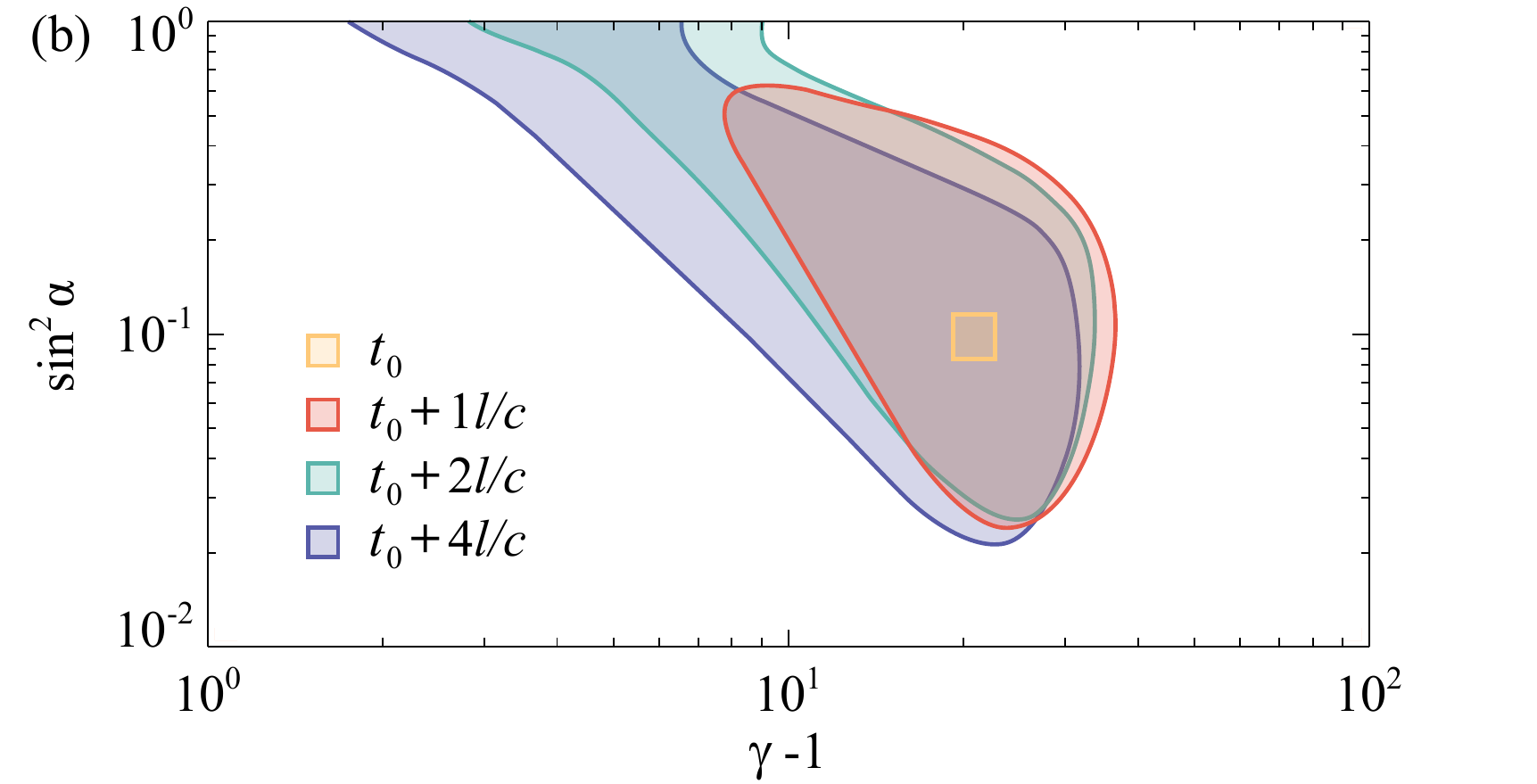}
\end{center}
\vspace{-0.4cm}
\caption{(a) $\langle \sin^2 \alpha \rangle$ as a function of the particle kinetic energy evaluated in the local ${\bm{E}} \times {\bm{B}}$ frame,  at $ct/l=2.5$ (solid lines) and $ct/l=10$ (dotted-dashed lines) for simulations with $\gamma_{\rm{rad}} \in \left\{ {75,125,200,\infty} \right\}$.  (b) From the $\gamma_{\rm{rad}}=75$ simulation, contours of the probability density functions (at 1 standard deviation) for particles starting in a selected $(\gamma-1,\sin^2 \alpha)$ bin at $t_0 = 3 l/c$ (yellow),  and followed after $c \Delta t/l = 1$ (red), $c \Delta t/l = 2$ (green), and $c \Delta t/l = 4$ (blue).}
\label{fig3}
\end{figure}

In Fig.~\ref{fig1}(a),  we show the cell-averaged mean particle Lorentz factor,  $\langle \gamma \rangle_{\rm{cell}}$,  from the most strongly cooled simulation ($\gamma_{\rm{rad}} = 75$).  Particle acceleration is highly localized in space,  which is a natural outcome of particle acceleration at fast-reconnecting current sheets  \cite{ComissoSironi18,ComissoSironi19} that form at different scales within the turbulence inertial range. Fig.~\ref{fig1}(b) shows that the turbulent cascade exhibits an extended inertial range with a magnetic power spectrum that approximately follows $P_B(k_\bot) \propto k_\bot^{-5/3}$ \cite{GS95,ThompsonBlaes98},  where $k_\bot = {(k_x^2 + k_y^2)^{1/2}}$,  and a slightly shallower electric power spectrum $P_E(k_\bot)$.  At kinetic scales ($k_\bot d_{e0} \gtrsim 0.5$),  both spectra steepen and approach $P_{B,E}(k_\bot) \propto k_\bot^{-6}$ (see also \cite{Zhda2020}).

A key outcome of the turbulent cascade is the generation of a nonthermal particle spectrum with a hard power-law range,  as shown in Fig. ~\ref{fig2}.  Figure~\ref{fig2}(a) presents the time evolution of the particle energy spectrum $dN/d\gamma$ for $\gamma_{\rm{rad}} = 75$.  Since particle injection via magnetic reconnection \cite{ComissoSironi18,ComissoSironi19} occurs on a timescale $t_{\rm{acc}} \sim \gamma_\sigma/ \omega_L \beta_E \ll l/c$,  the initial emergence of the power law ${dN}/{d\gamma} \propto \gamma^{-p}$ with $p \sim 1$ for $ct/l \sim 2-3$ is essentially unaffected by radiative cooling. At the same time,  a significant fraction ($\sim 25\%$) of the total kinetic energy is carried by particles having $\gamma > \gamma_{\rm{rad}}$ \cite{note1}.  Then,  the subsequent evolution of the particle spectrum ($ct/l \gtrsim 3$) reveals that the injected population of nonthermal particles becomes even harder ($p<1$) as particles cool down.

While the power-law slope $p$ is not affected by radiative cooling at early times ($ct/l = 2.5$),  the spectrum of the cooled distributions at late times ($ct/l = 10$) is markedly different from the uncooled one ($\gamma_{\rm{rad}} = \infty$),  as illustrated in Fig.~\ref{fig2}(b).  For $\gamma_{\rm{rad}} = \infty$,  a steeper power law extends from $\gamma \sim 30$ up to $\gamma \sim  \gamma_{\rm{max}}$, as a result of stochastic Fermi acceleration \cite{ComissoSironi18,ComissoSironi19,Comisso20,NB2021,Zhda2018,Wong2020}.  In contrast,  for strong cooling ($\gamma_{\rm{rad}} = 75, 125$),  a harder power law (with $p \sim 0.5$) extends from $\gamma \sim 30$ down to $\gamma \sim  \gamma_{\rm{cool}} (\kappa=10)$. This hardening of the cooled spectrum is in striking contrast with the standard textbook relation ${dN}/{d\gamma} \propto \gamma^{-\max[2,p+1]}$ \citep[e.g.][]{RL79} based on the \emph{ansatz} of isotropy for the velocity distribution of the synchrotron-emitting particles.

Both effects --- the late-time hardening of the particle spectrum and the physics of particles exceeding $\gamma_{\rm{rad}}$ --- can be understood by realizing that particle acceleration and cooling drive a significant energy-dependent anisotropy of the particle pitch angle $\alpha$. We show it in Fig.~\ref{fig3}(a),  where we display the mean of $\sin^2 \alpha$ as a function of $\gamma-1$.  We measured $\alpha$ and $\gamma$ in the local ${\bm{E}} \times {\bm{B}}$ frame, as this yields a straightforward evaluation of the synchrotron losses.  We find that $\langle \sin^2 \alpha \rangle$ deviates significantly from the expected mean of an isotropic distribution, $\langle \sin^2 \alpha \rangle = 2/3$ (compare with dotted line).  In particular,  at late times, $\langle \sin^2 \alpha \rangle$ attains a minimum in correspondence with the high end of the cooled power law ($\gamma \sim 30$) and approximately follows $\langle \sin^2 \alpha \rangle \propto \gamma^{-1.3}$ at lower $\gamma$. 
From the analysis of the trajectories of a subsample of $\sim 10^7$ particles,  we find that $\sin \alpha$ generally increases as the particles cool down.  This is illustrated in Fig.~\ref{fig3}(b),  where we show the evolution of particles starting from a given $(\gamma-1,\sin^2 \alpha)$ bin in the high energy end of the nonthermal tail (see yellow square).  Contours of the probability density functions (at 1 standard deviation) for the tracked particles show indeed that $\sin \alpha$ increases during cooling ($c \Delta t/l = 1, 2, 4$).  This statistical outcome reflects the fact that synchrotron cooling is biased towards cooling down those particles that get deflected to higher $\sin \alpha$.

The energy-dependent pitch-angle anisotropy regulates the evolution of the particle energy spectrum in view of $\tau_{\rm{cool}} \propto (\gamma \sin^2 \alpha)^{-1}$. It is straightforward to show that if we take $\sin \alpha \propto \gamma^{q}$ in the range of interest of the power law,  $\gamma_{\rm{cool}}  < \gamma <  \gamma_\sigma$,  the cooled distribution of particles injected via reconnection turns into the power law ${dN}/{d\gamma} \propto  \gamma^{-2(1+q)}$, assuming mono-energetic injection.  Thus,  if $q < 0$,  the cooled particle spectrum becomes harder than the common expectation for isotropic particles.

\begin{figure}[]
\begin{center}
\includegraphics[width=8.75cm]{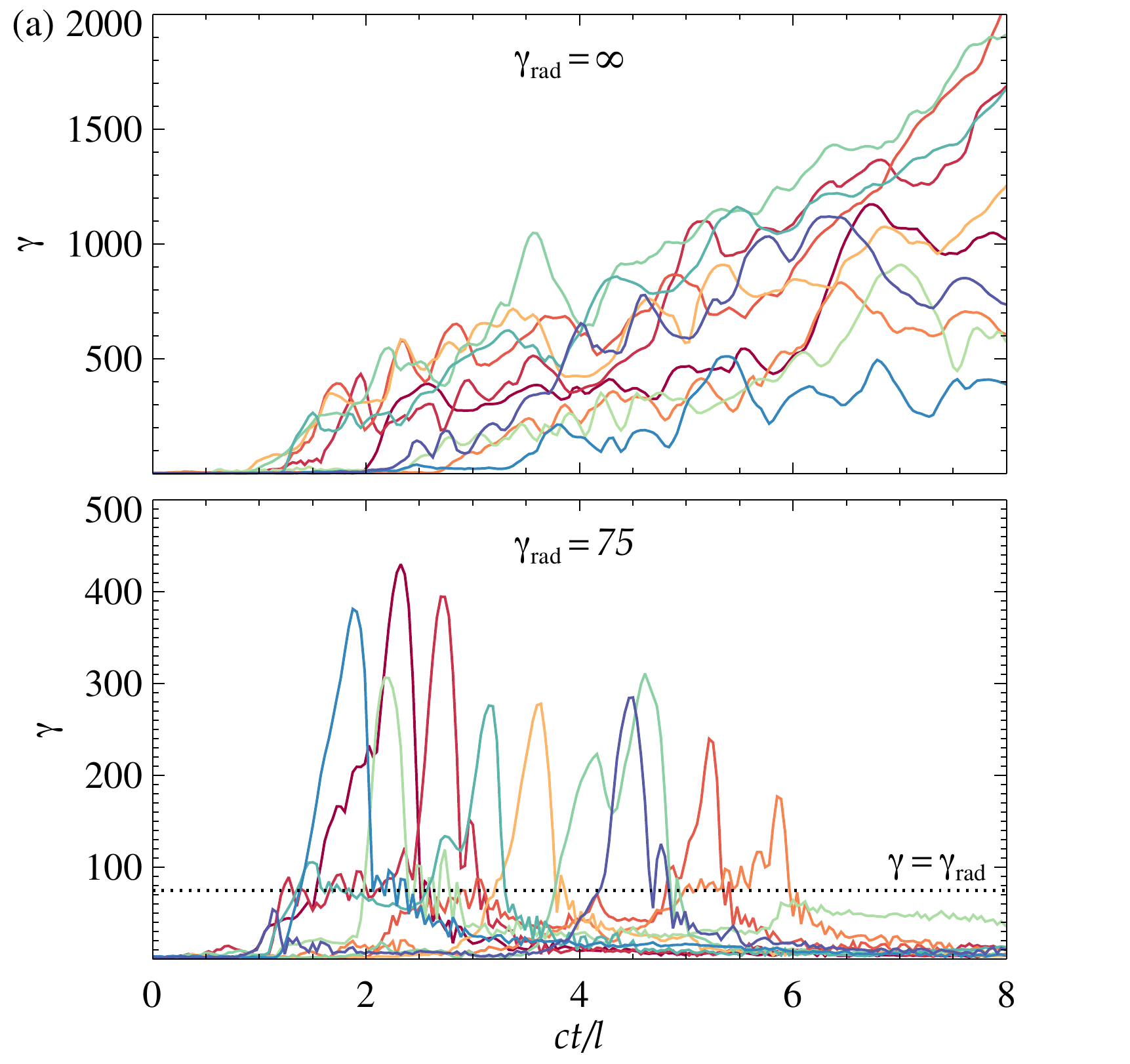}
\includegraphics[width=8.75cm]{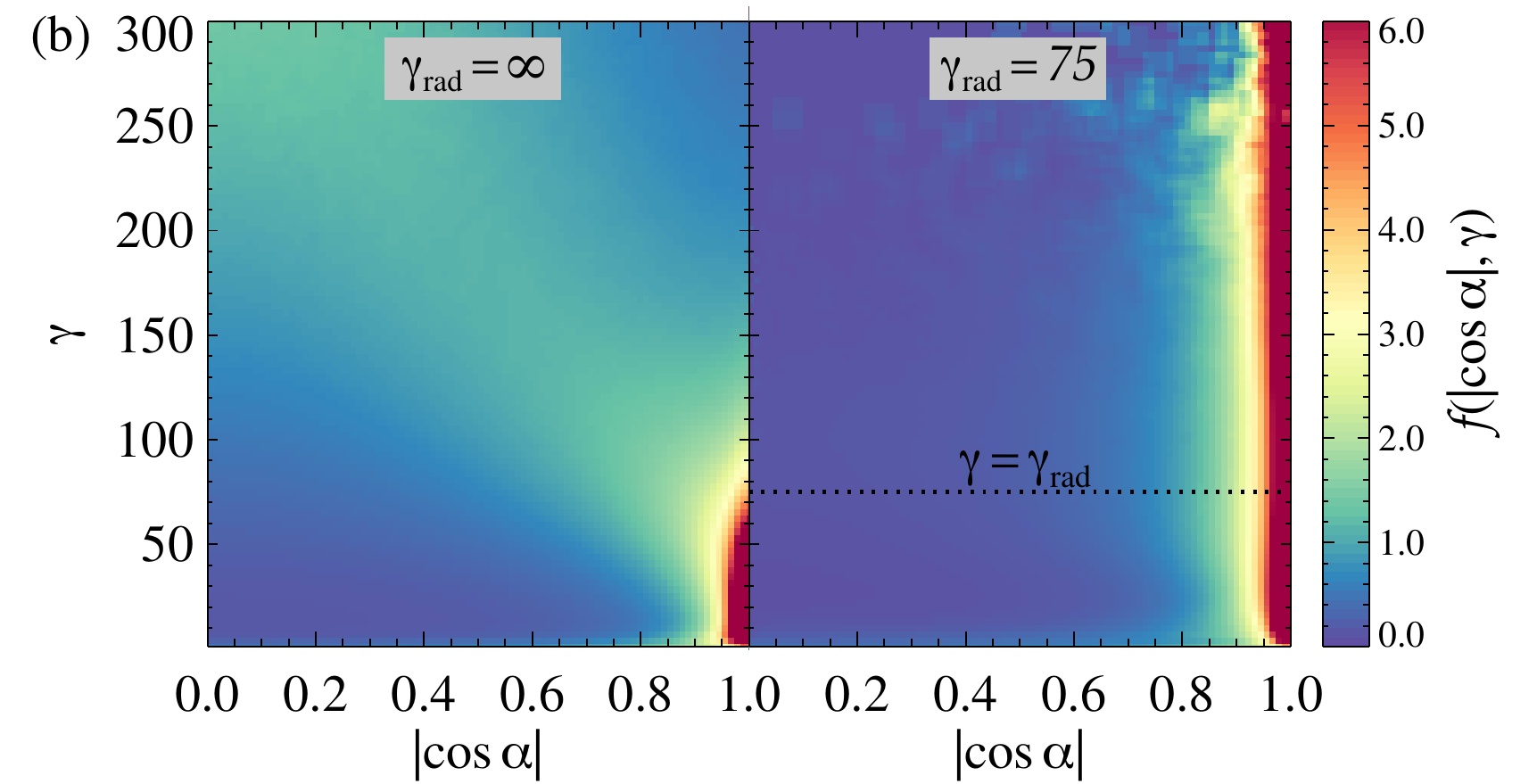}
\end{center}
\vspace{-0.4cm}
\caption{(a) Evolution of the Lorentz factor for 10 representative particles that attain $\gamma > 3 \gamma_\sigma$ during their evolution for $\gamma_{\rm{rad}}=\infty$ (top) and $\gamma_{\rm{rad}}=75$ (bottom).   (b) Particle distributions as a function of $|\cos \alpha|$ and $\gamma$ evaluated in the local ${\bm{E}} \times {\bm{B}}$ frame in the time range $2< ct/l < 6$ for $\gamma_{\rm{rad}}=\infty$ (left) and $\gamma_{\rm{rad}}=75$ (right).}
\label{fig4}
\end{figure}

Pitch-angle anisotropy also explains why some particles exceed $\gamma_{\rm{rad}}$. To illustrate this effect,  in Fig. \ref{fig4}(a) we show the Lorentz factor evolution of 10 representative particles that reach high energies for nonradiative (top panel) and radiative (bottom panel) turbulence.  In nonradiative turbulence,  stochastic acceleration following injection can drive particles up to $\gamma \sim \gamma_{\rm max}$ \cite{zhdankin17,ComissoSironi18,ComissoSironi19,Comisso20,NB2021,Zhda2018,Wong2020}.   Stochastic acceleration is suppressed in the fast-cooling regime,  but particles can still reach $\gamma \gg \gamma_{\rm{rad}}$ by being accelerated with $\sin \alpha \ll 1$, thus beating the otherwise rapid losses.  When particles reach $\gamma \gg \gamma_{\rm{rad}}$,  after a scattering event increases their pitch angle,  they abruptly radiate away most of their energy in a fraction of gyroperiod since $\tau_{\rm{cool}} (\gamma \gg \gamma_{\rm{rad}}) \ll  \gamma_{\rm{rad}}/(\omega_L \beta_E)$ for $\sin \alpha \sim 1$. Fig.~\ref{fig4}(b) shows the distribution $f \left( {|\cos \alpha|,\gamma} \right)$ with respect to $|\cos \alpha|$ and $\gamma$ evaluated in the local ${\bm{E}} \times {\bm{B}}$ frame, normalized such that ${\int_{-1}^1  {f\left( {|\cos \alpha|,\gamma} \right) d(|\cos \alpha|)}}  = 1$. While most high-energy particles have $\cos \alpha \sim 0$ in nonradiative turbulence (left panel), the distribution is strongly peaked at $|\cos \alpha| \sim 1$ in the radiative regime (right panel), as particles with $\cos \alpha \sim 0$ would necessarily be cooled down when $\gamma > \gamma_{\rm{rad}}$.

\begin{figure}
\begin{center}
\includegraphics[width=8.75cm]{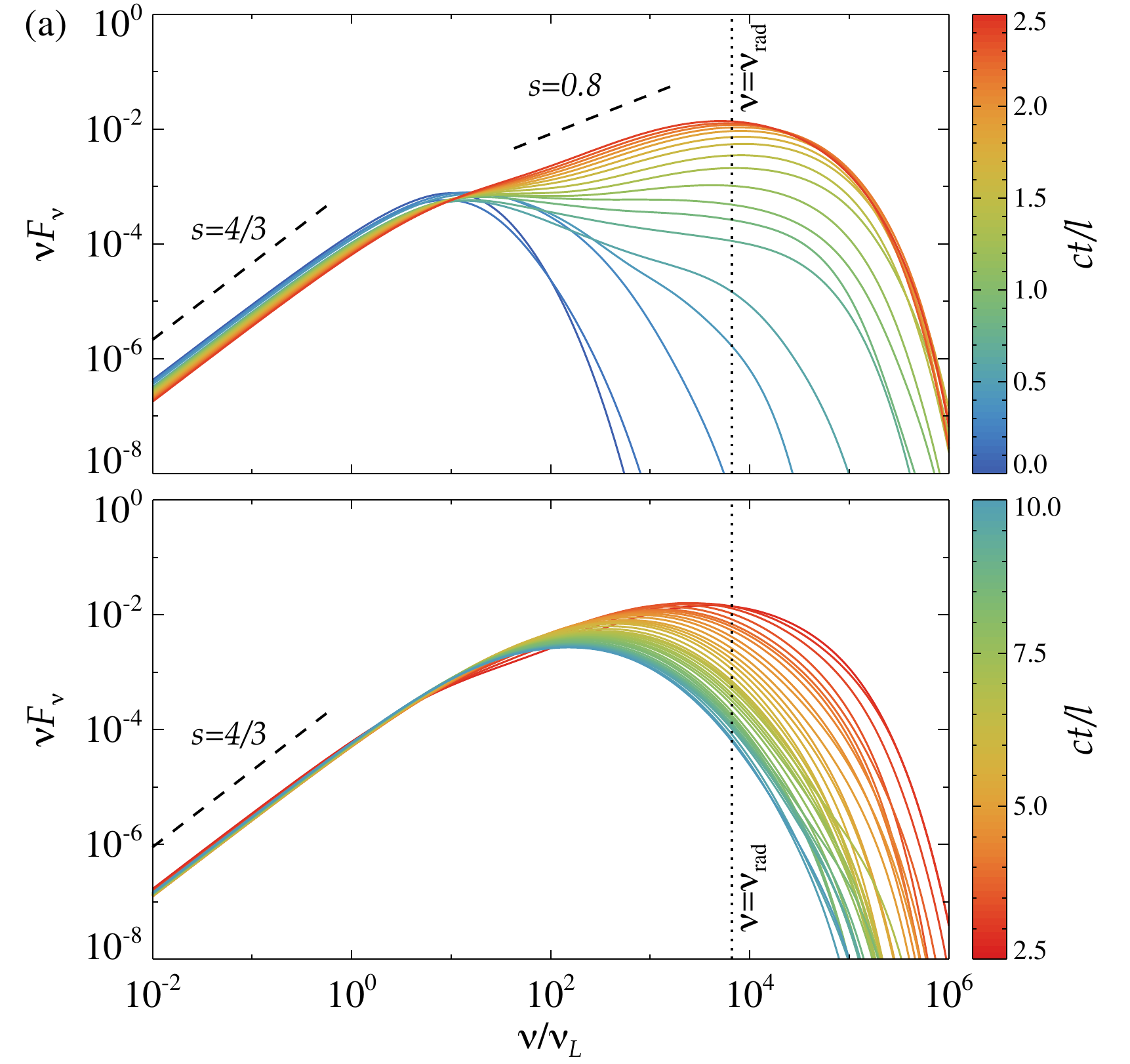}
\includegraphics[width=8.75cm]{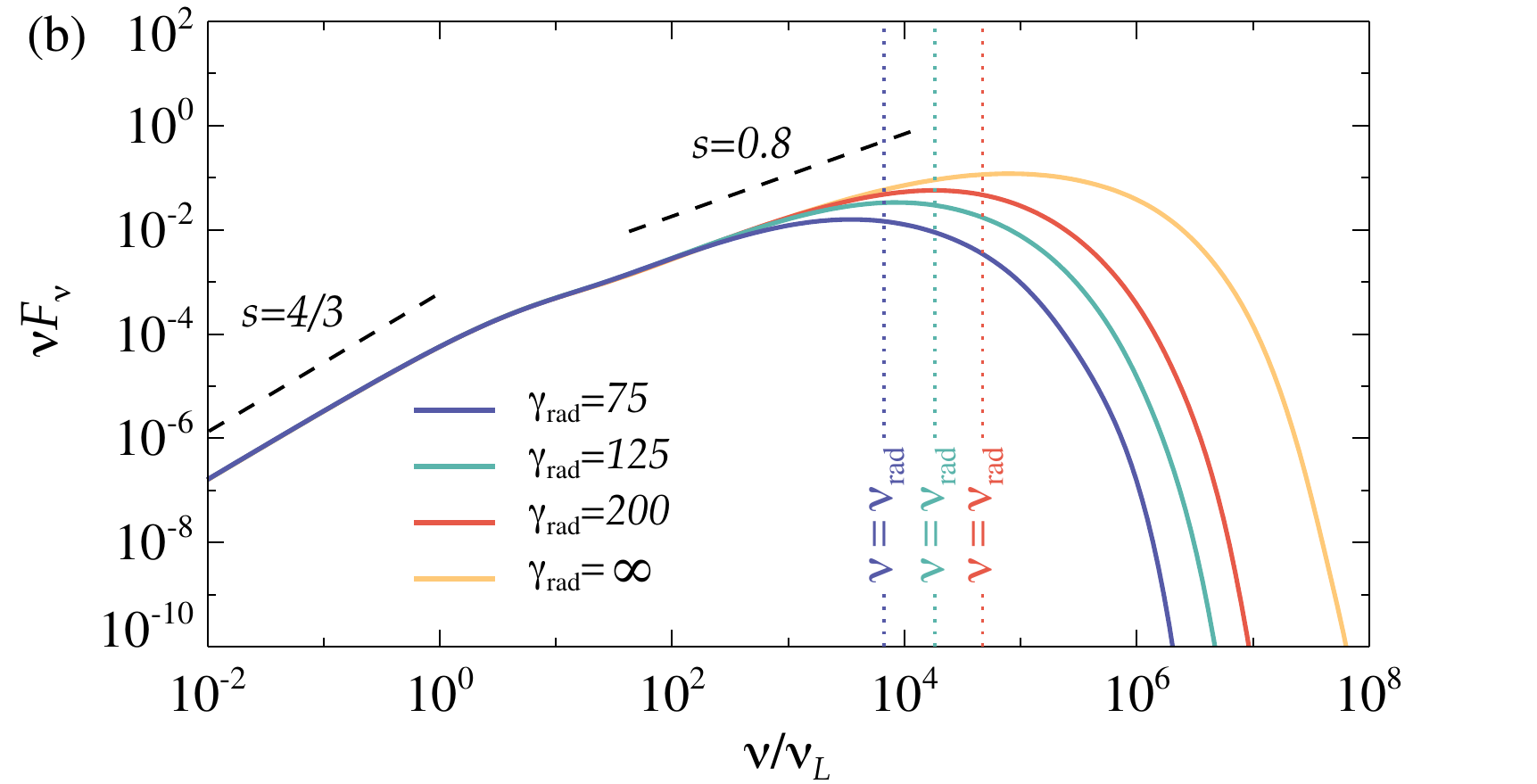}
\end{center}
\vspace{-0.4cm}
\caption{(a) Time evolution of the angle-integrated synchrotron spectrum separated in two time intervals for the simulation in Fig.~\ref{fig1}.  (b) Angle-integrated synchrotron spectra at peak luminosity ($ct/l=2.5$) for simulations with $\gamma_{\rm{rad}} \in \left\{ {75,125,200,\infty} \right\}$.  Vertical dotted lines indicate the corresponding (same color code) radiation-reaction-limited frequency. }
\label{fig5}
\end{figure}

We finally calculate the angle-integrated synchrotron spectrum \cite{RL79, RevilleKirk2010}.   Figure \ref{fig5}(a) shows the time evolution of the energy flux $\nu F_\nu$ for $\gamma_{\rm{rad}} = 75$. 
At low frequencies,  the synchrotron spectrum is the usual $\nu F_\nu \propto \nu^{4/3}$.  At higher frequencies,  we find $\nu F_\nu \propto \nu^{0.8}$ up to $\nu_{\rm{peak}} \sim \nu_{\rm{rad}} \sim \gamma_{\rm{rad}}^2  \nu_L$, where $\nu_L = \omega_L/2 \pi$.  A significant fraction ($\sim 35\%$) of radiative power is emitted above $\nu_{\rm{rad}}$ at $ct/l=2.5$ (the time of peak luminosity).  At later times, the high-energy end of the synchrotron spectrum recedes to lower frequencies as particles lose their energy and further particle acceleration fades, since most of the turbulent magnetic energy has already been dissipated.  Figure \ref{fig5}(b) shows the synchrotron spectrum at peak luminosity for different degrees of radiative cooling.  The spectral slope below $\nu_{\rm{peak}}$ is consistently hard and essentially unaffected by cooling.

In summary,  we have demonstrated that radiative relativistic plasma turbulence is a viable mechanism for fast particle acceleration and emission,  which self-consistently generates nonthermal particle spectra with a hard power-law range.  Particle acceleration can beat synchrotron losses by occurring at small pitch angles.  Consequently, particles can significantly exceed the nominal radiation-reaction limit before abruptly cooling down after pitch angle scattering.   
The anisotropy of the pitch-angle distribution, which is energy dependent,  also controls the evolution of the particle energy spectrum,  which is much harder than would otherwise be for an isotropic distribution.  
The turbulent energy cascade produces a power-law synchrotron spectrum $\nu F_\nu \propto \nu^s$  with $s \sim 1$ up to the synchrotron peak,  with a significant fraction of radiative power temporarily emitted above the nominal radiation reaction limit.  
These results have important implications for the origin of the nonthermal synchrotron emission from high-energy astrophysical sources,  in particular for the prompt phase of GRBs and gamma-ray flares from the Crab nebula.

\begin{acknowledgments}
We acknowledge fruitful discussions with Daniel Gro\v{s}elj, Joonas N\"attil\"a, and Emanuele Sobacchi.  This research acknowledges support from NASA 80NSSC20K1556,  NSF PHY-1903412,  DoE DE-SC0021254, and the Cottrell Fellowship Award RCSA 26932,  partially funded by the National Science Foundation (CHE-2039044).  The simulations were performed on Columbia University (Ginsburg), NASA-HEC (Pleiades), and NERSC (Cori) resources. 
\end{acknowledgments}


\begin{thebibliography}{}
\bibitem{note1} Similarly to what was observed in simulations of magnetic reconnection with low guide field \cite{Cerutti13,Yuan16,Kagan16,Hakobyan19}.
\end{thebibliography}
\end{document}